\documentclass{PoS}

\newcommand{\BR}{{\cal B}}

\newcommand{\pip}{\pi^+}
\newcommand{\pim}{\pi^-}
\newcommand{\piz}{\pi^0}
\newcommand{\gam}{\gamma}
\newcommand{\etap}{\eta^{\prime}}

\newcommand{\psp}{\psi^{\prime}}
\newcommand{\psip}{\psi^{\prime}}

\newcommand{\jpsi}{J/\psi}

\newcommand{\pp}{\pi^+\pi^-}

\newcommand{\ar}{\to}
\newcommand{\beq}{\begin{equation}}
\newcommand{\eeq}{\end{equation}}
\newcommand{\bitm}{\begin{itemize}}
\newcommand{\eitm}{\end{itemize}}

\newcommand{\chicz}{\chi_{c0}}

\newcommand{\chict}{\chi_{c2}}

\title{Recent results from BESIII}

\ShortTitle{Recent results from BESIII}

\author{\speaker{Chengping Shen}\thanks{On behalf of the BESIII collaboration and supported by the U. S. Department of Energy under Contracts No.
DE-FG02-04ER41291. }\\
        University of Hawaii, Honolulu, Hawaii 96822, USA\\
        E-mail: \email{shencp@phys.hawaii.edu}}

\abstract{Using about $226\times 10^6$ $\jpsi$ events and $106\times
10^6$ $\psp$ events collected with the BESIII detector at the BEPCII
$e^+e^-$ collider, the Dalitz plot of $\etap \to \eta \pi^+ \pi^-$
decay is studied, the direct measurement of
$a^{0}_{0}(980)-f_{0}(980)$ mixing is performed, $X(2120)$ and
$X(2370)$ are observed in the $\pi^+\pi^-\eta^\prime$ invariant mass
spectrum, $\psip\to\gamma\pi^0$ and $\psip\to\gamma\eta$ are
observed for the first time, and the decays $\chi_{cJ}\to\gamma V$
($V=\phi$, $\rho^0$, $\omega$) are studied.}

\FullConference{The Xth Nicola Cabibbo International Conference on Heavy Quarks and Leptons,\\
        October 11-15, 2010\\
        Frascati (Rome) Italy}

\begin{document}

\section{BESIII and BEPCII}

BESIII/BEPCII~\cite{bepc2} is a major upgrade of the BESII
experiment at the BEPC accelerator~\cite{bepc1} for studies of
hadron spectroscopy and $\tau$-charm physics. The analyses reported
here use data samples of $226\times 10^6$ $\jpsi$ events and
$106\times 10^6$ $\psp$ events.

\section{Measurement of the Matrix Element for the Decay $\etap \to \eta \pp$}

Investigation of matrix elements for particle decays is of paramount
importance for obtaining deeper insight into the dynamics of the
processes and into the structure of particles. The hadronic decays
of the $\etap$ meson have been extremely valuable in studies devoted
to chiral theory~\cite{ch3}, the effect of the gluon
component~\cite{gluon}, and the possible nonet of light
scalars~\cite{nonet}.

The branching fraction of $\jpsi \to \gamma \etap$ is measured to be
$(4.82\pm0.03~({\rm stat})\pm0.25~({\rm sys}))\times 10^{-3}~\cite{bes3etap}, $
which is consistent with the PDG value~\cite{PDG} within 1.5
$\sigma$. The Dalitz plot distribution for the decay $\etap \to \eta
\pp$ is described by two variables:
$X=\frac{\sqrt{3}}{Q}(T_{\pi^+}-T_{\pi^-})$, $
Y=\frac{m_{\eta}+2m_{\pi}}{m_{\pi}}\frac{T_{\eta}}{Q}-1$, where
$T_{\pi, \eta}$ denote the kinetic energies of mesons in the $\etap$
rest frame and $Q=T_{\eta}+T_{\pip}+T_{\pim}$. The squared absolute
value of the decay amplitude is expanded around the center of the
corresponding Dalitz plot: $M^2=A(1+aY+bY^2+cX+dX^2)$, where $a,b,c$
and $d$ are real parameters and A is a normalization factor. This
parametrization is called the general decomposition. A second
parametrization is the linear one: $M^2=A(|1+\alpha Y|^2+cX+dX^2)$,
where $\alpha$ is a complex parameter.

Dalitz plot parameters are obtained by minimizing $\chi^2(N, a, b,
c, d)= \sum_{i}^{n_{bin}}\frac{(D_i - N M_i)^2}{\sigma_{i}^2}$. Here
the index $i$ enumerates cells in Dalitz plot (empty cells outside
the Dalitz plot boundaries are excluded), N is normalization factor,
$a, b, c$ and $d$ are the Dalitz plot parameters. The $M_i$ and
$D_i$ are the numbers of (weighted) entries in the $i$-th bin of the
two-dimensional histograms in the Dalitz variables for MC and for
the background-subtracted data, respectively. The statistical error
$\sigma$ includes background subtraction and MC statistical errors.
The MC histogram is obtained as follows: $
M_i=\sum_{j=1}^{N_{ev}}(1+aY_j+bY_j^2+cX_j+dX_j^2), $ for the
general decomposition parametrization, where the index $j$ is over
the generated events and $X_j$ and $Y_j$ are the generated true
values of Dalitz variables. Similarly for the linear
parametrization, $M_i=\sum_{j=1}^{N_{ev}}(|1+\alpha
Y_j|^2+cX_j+dX_j^2)$.

\begin{figure}[h]
\centering
\includegraphics[width=60mm, angle=-90]{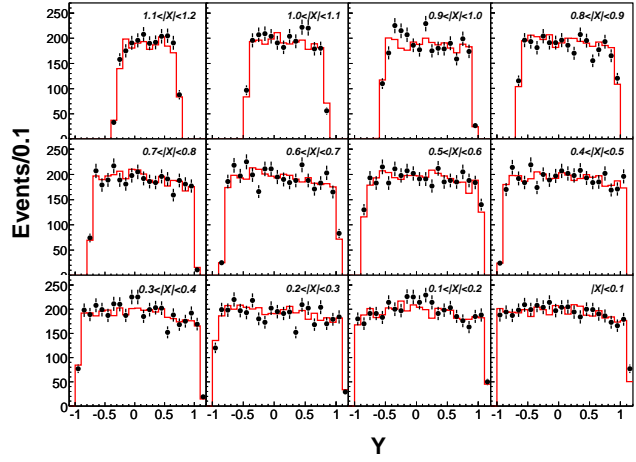}
\caption{Experimental distributions of the variable $Y$ in various
intervals of $X$ with the fitting function (histogram) for the
general decomposition parametrization.} \label{y-fit}
\end{figure}

The fitted values of the parameters of the matrix element for the
generalized and linear representations are: $
a=-0.047\pm0.011\pm0.003, b=-0.069\pm0.019\pm0.009, c=+0.019\pm
0.011\pm0.003, d=-0.073\pm0.012\pm0.003$, and $
\hbox{Re}(\alpha)=-0.033\pm0.005 \pm 0.003,
\hbox{Im}(\alpha)=0.000\pm0.049 \pm 0.001, c=+0.018\pm 0.009 \pm
0.003, d=-0.059\pm0.012 \pm 0.004$, respectively. Here the first
errors are statistical, the second systematic. The fitted results
are shown in Fig.~\ref{y-fit}.

\section{Study of $a^{0}_{0}$(980)--$f_{0}$(980) mixing}

The study of $a^{0}_{0}$(980) and $f_{0}$(980) nature is one of the
central topics in light hadron spectroscopy. The mixing between
$a^{0}_{0}(980)$ and $f_{0}(980)$ is expected to shed light on the
nature of these two resonances. Two kinds of mixing intensities
$\xi_{af}$ and $\xi_{fa}$ for the $a^{0}_{0}(980)\to f_{0}(980)$ and
$f_{0}(980)\to a^{0}_{0}(980)$ transitions are expressed
as~\cite{wujj2}: $\xi_{fa} = \frac {Br(J/\psi\to\phi
f_{0}(980)\to\phi a^{0}_{0}(980)\to\phi\eta\pi^{0})}
{Br(J/\psi\to\phi f_{0}\to\phi\pi\pi)}$, $ \xi_{af} = \frac
{Br(\psi'\to\gamma\chi_{c1}\to\gamma\pi^{0}
a^{0}_{0}(980)\to\gamma\pi^{0}
f_{0}(980)\to\gamma\pi^{0}\pi^{+}\pi^{-})}
{Br(\psi'\to\gamma\chi_{c1}\to\gamma\pi^{0}
a^{0}_{0}\to\gamma\pi^{0}\pi^{0}\eta)}$.

For the decay $J/\psi\to\phi f_{0}(980)\to\phi
a^{0}_{0}(980)\to\phi\eta\pi^{0}$, a simultaneous unbinned maximum
likelihood fit to the $\eta\pi^{0}$ mass spectra recoiling against
the $\phi$ signal and the $\phi$ mass sideband is performed. The fit
yields $N(f_{0}\to a^{0}_{0}) = 25.8\pm8.6$ (stat) events for the
mixing signal with a statistical significance of $3.4\sigma$. The
$\BR(J/\psi\to\phi f_{0}(980)\to\phi
a^{0}_{0}(980)\to\phi\eta\pi^{0})$ is measured to be
$(3.2\pm1.1$~(stat)$\pm0.8$~(sys)$)\times 10^{-6}$ ($<5.1\times
10^{-6}$ at 90$\%$ C.L.).

For the decay $\psi'\to\gamma\chi_{c1}\to\gamma\pi^{0}
a^{0}_{0}(980) \to\gamma\pi^{0}
f_{0}(980)\to\gamma\pi^{0}\pi^{+}\pi^{-}$, a simultaneous fit is
performed to the $\pi^{+}\pi^{-}$ invariant mass spectra in the
$\chi_{c1}$ mass window and the $\chi_{c1}$ mass sideband. The fit
yields $N(a^{0}_{0}\to f_{0})=6.4\pm3.2$ (stat) events for the
mixing signal with a statistical significance of $1.9\sigma$. The
$\BR(\psi'\to\gamma\chi_{c1}\to\gamma\pi^{0} a^{0}_{0}(980)
\to\gamma\pi^{0} f_{0}(980)\to\gamma\pi^{0}\pi^{+}\pi^{-})$ is
measured to be $(2.7\pm1.4$(stat)$\pm0.7$(sys)$)\times 10^{-7}$
($<5.9\times 10^{-7}$ at 90$\%$ C.L.).

The mixing intensity $\xi_{fa}$ for the $f_{0}(980)\to
a^{0}_{0}(980)$ transition is calculated to be
$\xi_{fa}=0.6\pm0.2~(\hbox {stat})\pm0.2~(\hbox {sys})\%$. The
mixing intensity $\xi_{af}$ for the $a^{0}_{0}(980)\to f_{0}(980)$
transition is calculated to be  $\xi_{af}
=0.3\pm0.2~(\hbox{stat})\pm0.1~(\hbox{sys})\%$.

\section{Observation of three resonances in
$\jpsi\to\gamma \etap \pp$}

The X(1835) is observed in the  $\jpsi\to\gamma \etap \pp$ with a
statistical significance of 7.7 $\sigma$ by the BESII experiment.
The possible interpretations of the X(1835) include a $p\bar{p}$
bound state, a glueball  and a radial excitation of $\etap$ meson,
etc. A high statistical data sample collected with the BESIII
provides an opportunity to confirm the existence of the X(1835) and
look for possible related states that decay to $\pp \eta'$.

\begin{figure}[h]
\centering
\includegraphics[width=60mm, height=5cm]{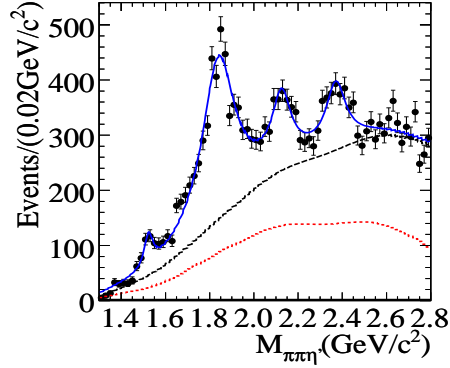}
\caption{The $\etap \pp$ invariant mass distribution for the
selected events from the two $\etap$ decay modes. The dash-dot line
is contributions of non-$\etap$ events and the $\etap \pp \pi^0$
background and the dash line is the total background contribution.}
\label{x1835}
\end{figure}

The $\etap \pp$ invariant mass spectrum for the combined two $\etap$
decay, $\etap \to \eta \pp$ and $\etap \to \gamma \rho$, is
presented in Fig.~\ref{x1835}. Two new resonances, the X(2120) and
the X(2370), are observed with statistical significances larger than
7.2$\sigma$ and 6.4$\sigma$, respectively. The masses and widths are
measured to be $M=1836.5\pm3.0^{+5.6}_{-2.1}$ MeV/$c^2$ and $\Gamma=
190 \pm 9^{+38}_{-36}$ MeV/$c^2$ for the X(1835), $M= 2122.4\pm
6.7^{+4.7}_{2.7}$ MeV/$c^2$ and $\Gamma= 83 \pm 16^{+31}_{-11}$
MeV/$c^2$ for the X(2120), $M=2376.3\pm8.7^{+3.2}_{-4.3}$ MeV/$c^2$
and $\Gamma=83\pm17^{+44}_{-6}$ MeV/$c^2$ for the X(2370),
respectively. The product branching fraction is
$\BR(J/\psi\rightarrow\gamma X(1835))
\BR(X(1835)\rightarrow\pi^{+}\pi^{-}\eta^\prime) = (2.71\pm0.09~{\rm
(stat)}^{+0.49}_{-0.35}~{\rm (syst)})\times10^{-4}$, and the
corresponding angular distribution of the radiative photon is
consistent with a pseudoscalar assignment for the X(1835). The mass
of the X(1835) is consistent with the BESII result, but the width is
significantly larger.

\section{Evidence for $\psip$ decays into $\gamma \pi^0$ and $\gamma\eta$}

The study of vector charmonium radiative decay to a pseudoscalar
meson $P=(\pi^0, \eta, \eta^{\prime})$ provides important tests for
various phenomenological mechanisms, such as vector meson dominance
model (VDM)~\cite{cz-report,korner-1983}, two-gluon couplings to
$q\bar{q}$ states~\cite{korner-1983}, mixing of
$\eta_c-\eta^{(\prime)}$~\cite{chao-1990}, and final-state radiation
by light quarks~\cite{cz-report}.

For $P=\eta$ or $\eta^\prime$, the ratio of $R_{J/\psi}\equiv
\mathcal{B}(J/\psi \to \gamma \eta)/\mathcal{B}(J/\psi \to \gamma
\eta^\prime)$ can be predicted by first order of perturbation
theory~\cite{cz-report}, and comparing to $\eta$ and $\eta^\prime$
production in $\psip$ radiative decays, the same ratio $R_{\psip}$
can be defined and $R_{\psip} \approx R_{J/\psi}$ is
expected~\cite{cleo-c-2009-gp}. The decay $\psip \to \gamma \pi^0$
is suppressed because the photon can only be from final state
radiation off one of the quarks.  In Ref.~\cite{rosner_prd_2010},
the contribution from $\psip \to \gamma^* \to \gamma \pi^0$ is
calculated and $\mathcal{B}(\psip \rightarrow \gamma
\pi^0)=2.19\times 10^{-7}$ is obtained, which is compatible to the
VDM contribution and does not contradict the upper limit $5.0\times
10^{-6}$ (at 90\% C.L.) reported by the CLEO
Collaboration~\cite{cleo-c-2009-gp}.

The branching fractions are listed in
Table~\ref{final_br}~\cite{zhanglei}. We obtain
$R_{\psip}=(1.10\pm0.38\pm0.07)\%$, which is the first measurement
and is below the 90\% C.L. upper bound determined by the CLEO
Collaboration~\cite{cleo-c-2009-gp}. The corresponding $\eta :
\etap$ production ratio at $J/\psi$ resonance was measured to be
$R_{J/\psi}=(21.1\pm0.9)\%$~\cite{cleo-c-2009-gp}. $R_{\psip}$ is
unexpectedly smaller than that at the $\jpsi$ resonance by an order
of magnitude. Such a small value of $R_{\psip}$ poses a great
challenge to our understanding of the decay properties of the
charmonium states.

\newcommand{\rb}[1]{\raisebox{2.0ex}[0pt]{#1}}
\begin{table}[hbtp]
  \footnotesize
  \vspace{-0.4cm}
  \label{final_br}
  \begin{center}
     \renewcommand{\arraystretch}{1.1}
     \begin{tabular}{l|ccc}
        \hline
        Mode         &   BESIII  & Combined BESIII  & PDG  \\
        \hline

        $\psip\ar\gam\piz$   & $1.58\pm0.40\pm0.13$ & $1.58\pm0.40\pm0.13$ & $\leq5$ \\

        $\psip\ar\gam\eta(\pip\pim\piz)$  & $1.78\pm0.72\pm0.17$ &                      &         \\

        ~~$\ar\gam\eta(\piz\piz\piz)$     & $1.07\pm0.65\pm0.08$ & \multicolumn{1}{c}{\rb{$1.38\pm0.48\pm0.09$}} & \multicolumn{1}{c}{\rb{$\leq2$}}\\

        $\psip\ar\gam\etap(\pip\pim\eta)$ & $120\pm5\pm8$        &                      &         \\

        ~~$\ar\gam\etap(\pip\pim\gam)$    & $129\pm3\pm8$        & \multicolumn{1}{c}{\rb{$126\pm3\pm8$}} & \multicolumn{1}{c}{\rb{$121\pm8$}}\\

        \hline

      \end{tabular}
      \caption{Branching fractions ($10^{-6}$). The first errors are statistical
  and the second systematic.}
      \vspace{-0.5cm}
  \end{center}
\end{table}




\section{Study of $\chi_{cJ}$ radiative decays into a vector meson}

$J/\psi$ and $\psip$ double radiative decays $\psi\to\gamma X\to
\gamma\gamma V$ ($\rho^0$, $\omega$, $\phi$) provide a favorable
place to extract information on the flavor content of the $C$-even
resonance $X$ and to study gluon hadronization dynamics. The recent
CLEO experimental results~\cite{cleoc-ggv} for ${\cal
B}(\chi_{c1}\to \gamma \rho^0$, $\gamma\omega$, and $\gamma\phi)$
are an order of magnitude higher than the corresponding theoretical
predictions~\cite{zhaogd}. By including hadronic loop contributions,
a recent perturbative Quantum ChromoDynamics (pQCD)
calculation~\cite{chen} obtains results in agreement with the
experimental measurements of ${\cal B}(\chi_{c1}\to\gamma V)$.

After the event selection criteria, we can see clear $\chi_{c1}$
signals in all decay modes, while $\chicz$ and $\chict$ signals are
insignificant. Each of the distributions is fitted with vector meson
mass sideband background plus a 2nd order polynomial background and
three $\chi_{cJ}$ resonances as the signal shapes. The measured
branching fractions are summarized in Table~\ref{tabtotal}.

\begin{table}
\footnotesize
\begin{center}
\begin{tabular}{lc|lc|lc }\hline
 Mode  & $\BR$ ($\times 10^{-6}$)  &  Channel  & $\BR$ ($\times 10^{-6}$) & Channel  & $\BR$ ($\times 10^{-6}$)\\
  \hline
 $\chi_{c0}\to\gamma\phi$&$9.5\pm4.2\pm0.8$ &$\chi_{c0}\to\gamma\rho^0$&$<$10.2& $\chi_{c0}\to\gamma\omega$&  $<$12.7\\
 $\chi_{c1}\to\gamma\phi$&$25.8\pm5.2\pm2.0$ & $\chi_{c1}\to\gamma\rho^0$ & $228\pm13
 \pm16$ & $\chi_{c1}\to\gamma\omega$&$69.7\pm7.2\pm5.6$ \\
 $\chi_{c2}\to\gamma\phi$&$<$8.0& $\chi_{c2}\to\gamma\rho^0$  & $<$20.4
 & $\chi_{c2}\to\gamma\omega$&$<$6.0 \\\hline
\end{tabular}
 \caption{\label{tabtotal}Results of $\chi_{cJ}\to
\gamma V$. The upper limits are at the 90\% C.L.}
\end{center}
\end{table}

The longitudinal (transverse) polarization exhibits a
$\cos^2\Theta~(\sin^2\Theta)$ dependence~\cite{cos}, and the angular
distribution is expressed as: $\frac{dN}{d \cos\theta} \propto
|A_L|^2\cos^2\Theta +
 \frac{1}{2}|A_T|^2 \sin^2\Theta$, where $A_L$ and $A_T$ are the
 longitudinal and transverse polarization amplitudes, respectively, and $\Theta$ is
 defined as the angle between the vector meson flight direction in the
 $\chi_{cJ}$ rest frame and either the $\pi^+/K^+$ direction in the
 $\rho^0/\phi$ rest frame or the normal to the $\omega$ decay plane in
 the $\omega$ rest frame.

By fitting the angular distributions, the values of the fraction of
the transverse component are determined to be
$0.29_{-0.12-0.09}^{+0.13+0.10}$, $0.158\pm0.034^{+0.015}_{-0.014}$,
and $0.247_{-0.087-0.026}^{+0.090+0.044}$ for
$\chi_{c1}\to\gamma\phi$, $\chi_{c1}\to\gamma\rho^0$, and
$\chi_{c1}\to\gamma\omega$, respectively.
Our measurement of the polarization of the vector mesons indicates
that the longitudinal component is dominant in $\chi_{c1}\to\gamma
V$ decays, as expected for an axial-vector particle decaying into a
vector ($\phi$, $\rho^0$, and $\omega$) and a $\gamma$ in the
framework of the vector dominance model.

\section {Summary}

Using about $226\times 10^6$ $\jpsi$ events and $106\times 10^6$
$\psp$ events, some recent preliminary results are shown here. They
include the analysis of Dalitz plot of $\etap \to \eta \pi^+ \pi^-$
decay, the direct measurement of $a^{0}_{0}(980)-f_{0}(980)$ mixing,
the observations of two new resonances, $X(2120)$ and $X(2370)$, in
the $\pi^+\pi^-\eta^\prime$ invariant mass spectrum, the studies of
$\psip\to\gamma\pi^0$, $\psip\to\gamma\eta$, and $\chi_{cJ}\to\gamma
V$ ($V=\phi$, $\rho^0$, $\omega$).


\end{document}